\documentclass[3p,review]{elsarticle}

\usepackage{graphics}
\usepackage{lineno}

\usepackage{amsmath,amssymb,amsfonts,amsthm,amsbsy,amstext,stmaryrd}
\usepackage{siunitx}		
\usepackage{multirow} 	
\usepackage{booktabs} 	

\usepackage[usenames,dvipsnames,table]{xcolor}

\biboptions{square,comma,sort&compress}

\journal{Journal Of Nuclear Materials}

\begin{document}

\begin{frontmatter}

\title{Hydrogen solubility in zirconium intermetallic second phase particles.}

\author[IC,ANSTO]{P.A. Burr}
\author[IC]{S.T. Murphy}
\author[IC,HMS]{S.C. Lumley}
\author[IC]{M.R. Wenman\corref{Wenman}}
\author[IC]{R.W. Grimes}
\address[IC]{Centre for Nuclear Engineering and Department of Materials, Imperial College London, London, SW7 2AZ, UK.}
\address[ANSTO]{Institute of Materials Engineering, Australian Nuclear Science \& Technology Organisation, Menai, New South Wales 2234, Australia.}
\address[HMS]{Nuclear Department, Defence Academy, HMS Sultan, Gosport, Hampshire, PO12 3BY, UK.}
\cortext[Wenman]{Corresponding author. Tel: +44 (0)2075946763\\ \indent \indent E-mail address: m.wenman@imperial.ac.uk}

\begin{abstract}
The enthalpies of solution of H in Zr binary intermetallic compounds formed with Cu, Cr, Fe, Mo, Ni, Nb, Sn and V were calculated by means of density functional theory simulations and compared to that of H in $\alpha$-Zr.
It is predicted that all Zr-rich phases (formed with Cu, Fe, Ni and Sn), and those phases formed with Nb and V, offer lower energy, more stable sites for H than $\alpha$-Zr.
Conversely, Mo and Cr containing phases do not provide preferential solution sites for H.
In all cases the most stable site for H are those that offer the highest coordination fraction of Zr atoms. Often these are four Zr tetrahedra but not always.
Implications with respect to H-trapping properties of commonly observed ternary phases such as Zr(Cr,Fe)$_2$, Zr$_2$(Fe,Ni) and Zr(Nb,Fe)$_2$ are also discussed.
\end{abstract}

\end{frontmatter}



\section{Introduction}
\label{sec:intro}
Zirconium alloys are widely used as fuel cladding and structural materials in water cooled nuclear reactors.
The principal role of the cladding is to provide a physical barrier between the fuel and the coolant thereby ensuring radionuclides remain contained within the fuel pin, while protecting the nuclear fuel from a strong flow of hot, potentially corrosive water or water/steam mix.

In the past four decades a number of compositional and processing changes have been made to improve the mechanical and corrosion resistance properties of Zr alloys\cite{Sabol2005}.
Concurrently with the increased durability and reliability of the cladding (and structural elements of the reactor core) the fuel burnup has also increased, leading to a reduction in energy production costs.

One of the limiting factors for a further increase in fuel burnup is the hydrogen pick-up of Zr alloys.
The presence of H in the cladding is of concern for a variety of reasons including dimensional changes, reduced ductility of the metal, and the formation of hard, brittle hydrides, which in turn may increase corrosion rates or cause failure by delayed hydride cracking\cite{Strasser2008}. New nuclear fuel cycle regulations that take into account hydrogen levels in the cladding are also being considered\cite{Raynaud2011}.
It is, therefore, essential to develop a deeper understanding of the processes responsible for the H pick-up, in order to deliver improved reactor performance, from both economic and safety standpoints.
Recent work\cite{Wang2013} has been carried out to investigate, from first principles, the structure, stability and mechanical properties of hydrides within the Zr metal. The current work, instead, focuses on interaction between H and the second phase particles (SPPs) formed within the Zr alloys.

Most alloying elements that are typically used in Zr alloys have very limited solubility in $\alpha$-Zr, and therefore tend to precipitate as intermetallic SPPs\cite{Lumley2013}.
The main exceptions are Sn, which exhibits solid solubility in $\alpha$-Zr at alloying concentrations of interest, and Nb, which precipitates out as $\beta$-(Zr,Nb) solid solution at operating temperature and below, as well as forming intermetallic SPPs in the presence of Fe. Further details are provided in Section~\ref{sec:xtal}.
The interaction between the SPPs and H is not yet well understood. It remains unclear whether the SPPs act as traps for H, provide nucleation sites for hydrides, or whether they may act as a preferred transport path through the outer oxide layer for ingress of H into the metal\cite{Cox1999b,Hatano1996a}.

In our previous work\cite{Burr2013} we employed atomic scale computer simulations, based on density functional theory (DFT), to investigate the solubility of H in Zr intermetallics of particular importance for Zircaloy-2 and Zircaloy-4 alloys. Here, the study is extended to other binary intermetallics of Zr (containing Cu, Fe, Mo, Ni, Nb, Sn and V), which are either directly relevant to fuel cladding in water cooled nuclear reactors, or have been used in model alloys to understand the role SPPs play in controlling the H pick-up.

\section{Computational Methodology}
\label{sec:meth}
All DFT simulations were carried out using CASTEP\cite{Clark2005}.
The exchange and correlation functional employed was the generalized gradient approximation, as formulated by the Perdew Burke and Ernzerhof (PBE)\cite{Perdew1996}.

Ultra-soft pseudo potentials with a consistent cut-off energy of \SI{450}{\electronvolt} were used throughout.
A high density of {\bf k}-points was employed for the integration of the Brillouin zone, following the Monkhost-Pack sampling scheme\cite{Monkhorst1976}: the distance between sampling points was maintained as close as possible to \SI{0.30}{nm^{-1}} and never above \SI{0.35}{nm^{-1}}. In practice this means a sampling grid of $3\times3\times3$ points for the largest intermetallic supercells. 
The fast Fourier transform grid was set to be twice as dense as that of the wavefunctions, with a finer grid for augmentation charges scaled by 2.3.
Due to the metallic nature of the system, density mixing and 
Methfessel-Paxton\cite{Methfessel1989} cold smearing of bands were employed with a width of \SI{0.1}{eV}. Testing was carried out to ensure a convergence of \SI{e-3}{eV/atom} was achieved with respect to all of the above parameters. No symmetry operations were enforced and all calculations were spin polarised, taking particular care in finding the lowest energy spin state of phases containing magnetic elements.

The energy convergence criterion for self-consistent calculations was set to \SI{1e-6}{\electronvolt}. 
Similarly robust criteria were imposed for atomic relaxation: energy difference $<$~\SI{1e-5}{\electronvolt}, forces on individual atoms $<$~\SI{ 0.01}{\electronvolt\per\angstrom} and for constant pressure simulations stress component on cell $<$~\SI{0.05}{\giga\pascal}.

\section{Crystallography of intermetallic phases}
\label{sec:xtal}

Many different types of SPPs have been reported in the literature relating to Zr alloys, often with little agreement on their chemical composition and crystal structure.
In this section we report a summary of the possible SPPs that can form between Zr and each of the alloying elements under investigation (see Table~\ref{tab:SuperCells}).

\begin{table}[hbt]
\centering
\caption{ \label{tab:SuperCells} Overview of all compounds investigated for the accommodation of H, together with the size of the largest supercell simulated in terms of the number of atoms ($N$) and smallest distance between the H defect and its replicas ($d$). For Laves phases, M = Cr, Fe, Mo, Nb, V.}
\begin{tabular}{l l c c c c c}
\toprule
Phase			&Space	&Pearson		&Prototype	&{\it N}	&\it{d} &[H]\\
				&Group	&Symbol	&&&\it{\si{(nm)}}	& {\it wt}~ppm\\
\midrule
$\alpha$-Zr		&$P6_{3}/mmc$	&$hP2$	&Mg			&150	&1.56	&74\\
$\beta$-Zr,Nb		&$Im\bar3m$	&$cI2$	&W			&128	&1.43	&85--86\\
\midrule
ZrM$_2$ {\it C15}	&$Fd\bar3m$	&$cF24$	&Cu$_2$Mg	&192	&1.41	&57--81\\
ZrM$_2$ {\it C14}	&$P6_{3}/mmc$	&$hP12$	&MgZn$_2$	&96	&1.00	&114--163\\
ZrM$_2$ {\it C36}	&$P6_{3}/mmc$	&$hP24$	&MgNi$_2$	&96	&1.00	&114--163\\
Zr$_3$Fe			&$Cmcm$		&$oS16$	&Re$_4$B		&96	&1.14	&127\\
Zr$_2$(Fe,Ni)		&$I4/mcm$	&$tI12$	&Al$_2$Cu		&96	&0.98	&131--132\\
ZrNi				&$Cmcm$		&$oS8$	&TlI			&96	&1.00	&140\\
Zr$_2$Cu			&$I4/mmm$	&$tI6$	&MoSi$_2$		&96	&1.12	&128\\
ZrCu				&$Pm\bar3m$	&$cP2$	&CsCl	 	&128	&1.31	&102\\
Zr$_3$Sn			&$Pm\bar3n$	&$cP8$	&Cr$_3$Si		&64	&1.13	&160\\
Zr$_5$Sn$_{3}$		&$P6_{3}/mcm$	&$hP16$	&Mn$_5$Si$_3$	&128	&1.16	&78\\
Zr$_5$Sn$_{3.5}$	&$P6_{3}/mcm$	&$hP17$	&			&136	&1.19	&74\\
Zr$_5$Sn$_{4}$		&$P6_{3}/mcm$	&$hP18$	&Ga$_4$Ti$_5$	&143	&1.20	&68\\
ZrSn$_2$			&$Fddd$		&$oF24$	&TiSi$_2$		&288	&1.71	&32\\
\bottomrule
\end{tabular}
\end{table}

Most alloying additions tend to form intermetallic phases with Zr. Exceptions are Sn, which shows large solubility in $\alpha$-Zr and Nb, which is a $\beta$ phase stabiliser in Zr and as such, tends to form precipitates of $\beta$-(Nb,Zr) solid solutions within the $\alpha$-Zr matrix, as well as the intermetallic Zr(Nb,Fe)$_2$ in the presence of Fe.
Limitations in the current methodology do not allow the modelling of solid solutions, therefore the HCP and BCC phases of the pure elements were modelled instead. 

The most common intermetallic phases observed are the ZrM$_2$ Laves phases, where M = Cr, Fe, Mo, Nb, V (see Table~\ref{tab:SuperCells}).
Among the literature many reports exist of both the cubic C15\cite{Barberis2004,Yang1987,Vitikainen1978,Krasevec1981,Kuwae1983a,Versaci1979} and the two hexagonal C14 and C36\cite{Barberis2004,Krasevec1981,Kuwae1983a,Chemelle1983,Lelievre2002,Yang1986,Vandersande1974,Versaci1979,Versaci1983} Laves phases.
Due to the nominal composition of commercial alloys, these SPPs have mostly been identified as Zr(Cr,Fe)$_2$ and Zr(Nb,Fe)$_2$, however, they are also known to form with Mo and V additions\cite{Toffolon-masclet2002,Barberis2004,Shishov2013,Zinkevich2002,Stein2005}.

The Ni-Zr binary phase diagram exhibits numerous intermetallic compounds\cite{Okamoto2007}. However, Ni containing SPPs in common Zr alloys tend to be stable as Zr rich phases, typically the body-centred tetragonal Zr$_2$Ni phase, which --- in the presence of Fe --- forms Zr$_2$(Ni,Fe), as it is commonly found in Zircaloy-2\cite{Barberis2005,Chemelle1983}.
For completeness, here we also consider the orthorhombic ZrNi structure.

In addition to the ZrFe$_2$ and Zr$_2$Fe phases described above, the Zr-Fe system exhibits an orthorhombic Zr$_3$Fe phase, which has been observed only sporadically in high Fe, low Ni low Cr alloys\cite{Barberis2005}, $\beta$-quenched alloys\cite{Bangaru1985} and irradiated Zircaloys\cite{Griffiths1988}.
A metastable Zr$_4$Fe phase has also been reported\cite{Barberis2005,Nikulina1996,Yang1987,Yang1986,Griffiths1988}. Unfortunately there is insufficient crystallographic information to conduct a reliable DFT study of Zr$_4$Fe, therefore it was not considered further. Finally the $Fd\bar 3 m$ structure of Zr$_2$Fe was also considered, as documented by Buschow\cite{Buschow1981a}, but the formation energy of this phase was found to be \SI{0.20}{eV} greater than the body-centred Zr$_2$Fe phase described above and is therefore discarded from further investigation.

Copper additions in Zr-base alloys, tend to be observed mainly as a tetragonal Zr$_2$Cu phase\cite{Motta2007}, even though the complete phase diagram presents a large number of possible intermetallic compounds\cite{Okamoto2012,Arias1990}.
For completeness, all reported phases containing $\geq50\%$ Zr, were investigated. Table~\ref{tab:Zr-Cu} contains a list of the phases considered, with those that are widely regarded as stable being marked by an asterisk. Enthalpies of formation from standard state $(E_f^{\circ})$ and from solid solution $(E_{f}^{sol})$ are also reported in Table~\ref{tab:Zr-Cu}, following reaction~\ref{eq:formation} and \ref{eq:dissolution} respectively (Enthalpies of formation for the other possible SPP phases were reported previously~\cite{Lumley2013}).
\begin{align}
\label{eq:formation}
n\text{Zr} + \mathrm{Cu} &\rightarrow \mathrm{Zr}_n\mathrm{Cu}\\
\label{eq:dissolution}
x \mathrm{Zr_{149}Cu} &\rightarrow x \mathrm{Zr}_n\mathrm{Cu} + (149-n)x \mathrm{Zr}
\end{align}•
\begin{table}[hbt]
\centering
\caption{\label{tab:Zr-Cu}  List of Zr-Cu intermetallic phases modelled in the current work. The standard enthalpy of formation $E_f^{\circ}$ was calculated from $\alpha$-Zr and FCC-Cu metals. Whilst $E_f^{sol}$,was calculated from an isolated substitutional Cu atom in a 150 atoms cell of $\alpha$-Zr. The quoted enthalpies are per formula unit.}

\begin{tabular}{l l l l r@{.}l r@{.}l l}
\toprule
Formula		&Space		&Pearson	&Prototype		&\multicolumn{2}{c}{$E_f^{\circ}$} &\multicolumn{2}{c}{$E_f^{sol}$}	&Ref.\\
unit			&group		&symbol	&structure		&\multicolumn{2}{c}{(eV)}	&\multicolumn{2}{c}{(eV)}\\
\midrule
Zr$_3$Cu		&$P4/mmm$	&$tP4$	&CuInPt$_2$	&  0&29	&  0&00	&\cite{Karlsson1951}\\
Zr$_2$Cu$^{\ast}$&$I4/mmm$	&$tI6$	&MoSi$_2$		&-0&41	&-0&70	&\cite{Nevitt1962,Sviridova2004}\\
Zr$_2$Cu		&$Fd\bar3m$	&$cF24$	&AuBe$_5$		&-0&20	&-0&48	&\cite{Sviridova2004}\\
ZrCu$^{\ast}$	&$Pm\bar3m$	&$cP2$	&CsCl		&-0&21	&-0&50	&\cite{Carvalho1980,Zhalko-Titarenko1994}\\
ZrCu			&$P1 21/m1$	&$mP4$	&NiTi			&-0&25	&-0&54	&\cite{Zhalko-Titarenko1994}\\
ZrCu			&$Cm$		&$mS16$	&			&-0&26	&-0&55	&\cite{Zhalko-Titarenko1994}\\
\bottomrule
\end{tabular}
\end{table}
The Zr$_3$Cu phase was found to be thermodynamically unstable (positive formation energy). The tetragonal $I4/mmm$ structure of Zr$_2$Cu was found to possess a substantially more stable formation energy compared to the cubic $Fd\bar3m$ structure. On the other hand, there is little difference in the formation energies calculated for the ZrCu compounds, suggesting that all three phases are likely to form in the Cu-Zr alloy.
However, ZrCu is a high temperature phase, which decomposes in a eutectoid reaction at temperatures below \SI{715}{\celsius}.
DFT does not incorporate the effect of vibrational energy associated with temperature, therefore DFT results alone are generally not indicative of the stability of high temperature phases. This is especially true when the differences in formation energy are very small, as in the ZrCu phases.
Experimental investigation suggests that the CsCl structure is the most stable at high temperatures\cite{Arias1990}.
Following the above results, the body-centred Zr$_2$Cu and the CsCl structure of ZrCu were studied.

Finally, we examine Zr-Sn intermetallic compounds. Although Sn is highly soluble in $\alpha$-Zr, there have been reports of Zr-Sn intermetallic SPPs in irradiated Zircaloy samples\cite{Nikulina1996,Griffiths1987,Griffiths1988,Yang1987}, suggesting that they form due to radiation enhanced diffusion of the alloying elements. Recent work\cite{Lumley2013} confirms this from a thermodynamic viewpoint. However, redeposition of Sn during TEM sample preparation has also been suggested as a possible cause for the formation of Zr-Sn intermetallics\cite{Yang1986}.

Even ignoring the effects of irradiation and ternary alloying elements on the stability of Sn-SPPs,
the equilibrium binary Sn-Zr system is rather complex\cite{Okamoto2010,Kwon1990}. At levels of Sn greater than the $\alpha$-Zr solid solution regime, the stable phases are Zr$_4$Sn, Zr$_5$Sn$_{3+x}$ and ZrSn$_2$. With the exception of the latter, the other phases exhibit a high degree of disorder.
Extensive experimental work by Kwon and Corbett\cite{Kwon1990} subsequently modelled by Baykov {\it et al.}\cite{Baykov2006}~shows that at equilibrium Zr$_5$Sn$_{3+x}$ has a large concentration of self-interstitial Sn, up to a stoichiometry of Zr$_5$Sn$_4$. These studies also showed that Zr$_4$Sn is a Zr-substitutional structure, which is derived from Zr$_3$Sn with one fifth of the Sn sites occupied by Zr atoms, i.e. $\text{Zr}_3(\text{Sn}_{0.8}\text{Zr}_{0.2})$.
Due to limitations in our computational methodology, such complex phases cannot be modelled reliably, instead, their parent phases were simulated:
the cubic Zr$_3$Sn structure, face centered orthorhombic ZrSn$_2$, and the hexagonal Zr$_5$Sn$_3$, as well as two of its interstitial derivatives, an ordered form of Zr$_5$Sn$_{3.5}$ in which only the $2b$ Wyckoff sites were occupied, and Zr$_5$Sn$_4$, in which all the interstitial sites are occupied.

\section{Results and Discussion}

\subsection{Choice of Supercell size}

When simulating point defects in solids using the supercell approach, it is possible to do so under constant pressure or constant volume conditions.
In the former, which is more computationally expensive and replicates alloying conditions, both the crystal's lattice parameters and atomic positions within the supercells are subject to energy minimisation; as a result the size and shape of the supercell is allowed to react to any internal stress resulting from the incorporation of a defect. 
In the latter case, constraints are applied to the lattice so that the cell's shape and volume are fixed during internal energy minimisation. This method most closely replicates the dilute case, however, if the supercell size is insufficiently large, defect-defect interactions will cause a sufficient stress build-up on the cell, such that a significant non-physical contribution to the calculated energies is added, which we term the constrained cell energy contribution. 

\begin{figure}[p]
\centering
\includegraphics[width = 0.45\textwidth]{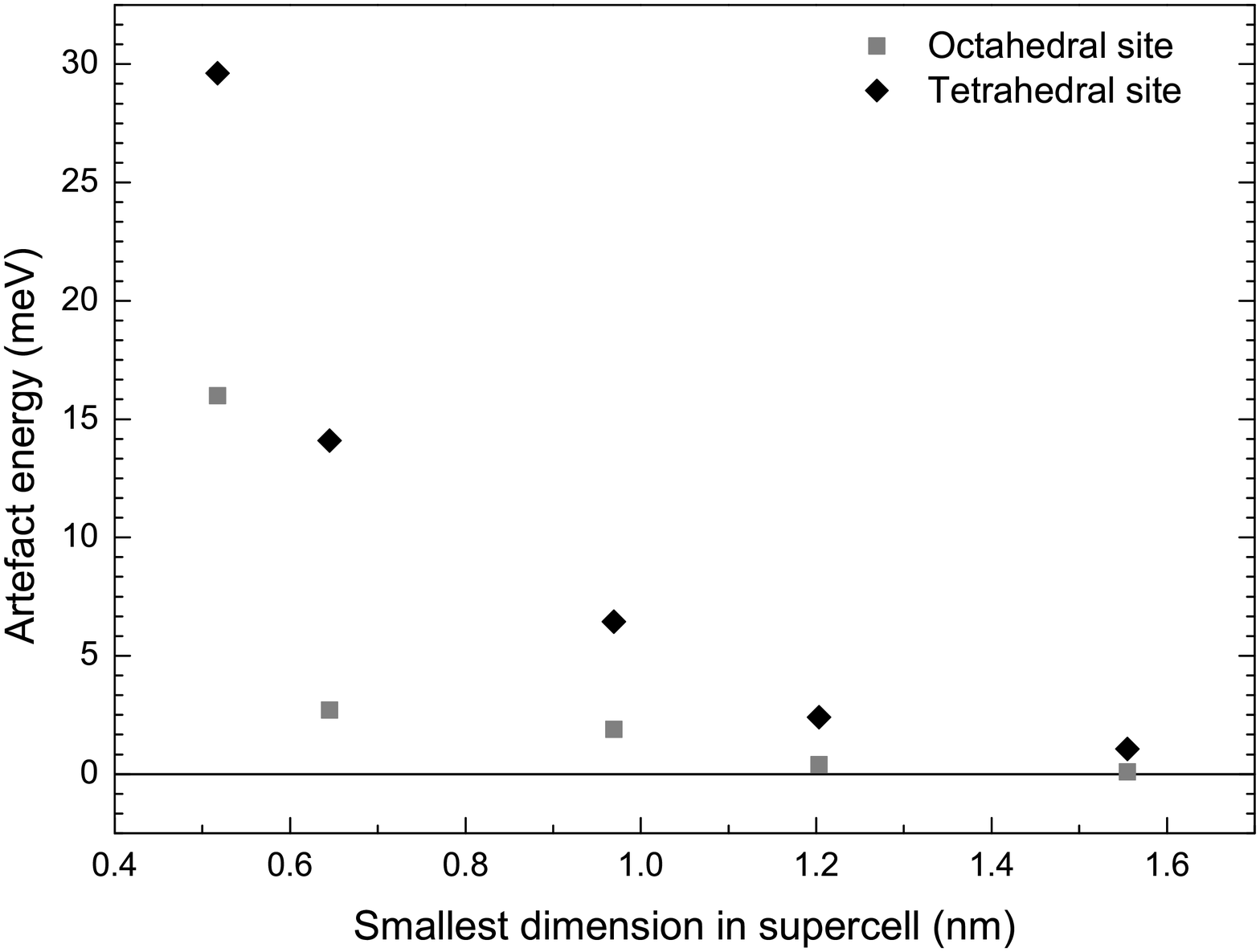}
\caption{\label{fig:aZr-SC-Conv} Constrained cell energy contribution (``energy of constant volume calculation'' $-$ ``energy of constant pressure calculation'') for an H interstitial defect in $\alpha$-Zr
vs.\ defect-defect separation, measured as the smallest dimension of the supercell (valid for $60^{\circ}<\alpha,\beta,\gamma<120^{\circ}$).}
\end{figure}

The difference in energy between the two methods is a good measure of how effectively the supercell describes isolated defects in the bulk material. 
A convergence analysis was carried out with respect to the supercell size for the H interstitial defects in $\alpha$-Zr, see Figure~\ref{fig:aZr-SC-Conv}.
It is clear that an accuracy of the order of $10^{-2}$ eV/atom is achieved with a defect-defect separation of \SI{0.8}{nm}, and $10^{-3}$ eV/atom at \SI{1.56}{nm} (corresponding to a $5\times5\times3$ supercell containing 150 Zr atoms).
Note that the tetrahedral interstitial has a larger defect volume, therefore it is affected by the size of the supercell to a greater extent.

Starting from a fully relaxed unit cell, two supercells were generated: first a smaller one, containing $\sim 50$ atoms, which was used for simulations of H defects in each of the crystallographically unique interstitial sites; then a larger one in which the lowest energy configurations of the defects, as identified from the smaller cell, were replicated for better accuracy. The larger supercells were chosen to have no supercell axis smaller than \SI{1}{nm} prior to relaxation. Before adding the defects to the supercells, these were relaxed again to avoid any aliasing errors (misalignment of atoms within the supercell where the crystal boundaries were situated in the original unit cell) and errors arising from the use of non-identical sampling grids.
This process was repeated for each intermetallic phase investigated.

\subsection{Hydrogen Accommodation}

As previously reported\cite{Burr2013}, H was found to preferentially occupy the tetrahedral site over the octahedral one in $\alpha$-Zr, and all other interstitial sites were found to be metastable. 
The enthalpy of solution of H in the tetrahedral site is -0.464 eV.
These findings are in agreement with both experimental results of Khoda-Bakhsh and Ross\cite{Khoda-Bakhsh1982} and the DFT results of Domain {\it et al.}\cite{Domain2002a}

Interestingly it was found that the lowest energy H solution site for nearly all intermetallic phases had a tetrahedral configuration. The only exceptions were ZrCu, ZrNi and Zr$_5$Sn$_{3+x}$, which are discussed in greater detail below. 
Furthermore, it was found systematically that sites with the largest fraction of neighbouring Zr atoms offered the lowest energy for H accommodation. Thus, for most of the intermetallic phases, the lowest energy site is one consisting of 4 neighbouring Zr atoms. For those compounds where, due to stoichiometry, no such sites are present, for example the ZrM$_2$ phases, H was found to preferentially occupy tetrahedral sites with 2--3 Zr atoms and only 1--2 M atoms, irrespective of the concentration of the M-species.
A recent study\cite{Gesari2010} reports an analogous behaviour for Laves phases.
In the case of ZrCu and ZrNi, the most stable interstices are octahedral with coordination of 4 Zr and 2 Cu/Ni atoms. These sites offer a greater Zr-neighbour fraction compared to the available tetrahedral sites, which have 2 Zr and 2 Cu/Ni neighbours.

A summary of solution enthalpies for H in the most favourable interstitial site, in each of the intermetallic phases, is presented in Table~\ref{tab:Esol_summary}.
\begin{table}[bt]
\centering
\caption{\label{tab:Esol_summary} Enthalpy of solution for an interstitial H, in the most stable site, for each of the intermetallic phases investigated. $\Delta E_{sol}^{\alpha\text{-Zr}}$(H) is the difference in enthalpy of solution of H in the given intermetallic, compared to the tetrahedral site in $\alpha$-Zr. The preferred interstitial site for H is indicated in Wyckoff notation, and it is described in terms of its geometry and coordination number with Zr and non-Zr elements X.
All values are expressed in units of eV. $^{\ast\ast} x=0\text{--}1$. $^\dag$From previous study\cite{Burr2013}.}
\begin{tabular}{l  r @{.} l r @{.} l r @{.} l c c r @{/} l}
\toprule
Phase & \multicolumn{2}{c}{$E_{sol}$(H)} & \multicolumn{2}{c}{$\Delta E_{sol}^{\alpha\text{-Zr}}$(H)}	&\multicolumn{2}{c}{error} & \multicolumn{4}{c}{H interstitial}\\ 
\multicolumn{7}{c}{ }&	Position	&Type	&Zr&X\\\midrule
$\alpha$-Zr		& -0&46	&\multicolumn{4}{c}{}				&$4f$	&tet	&4&--\\
$\beta$-Zr			& -0&62	&\bf-0&\bf16	&\multicolumn{2}{c}{}	&$12d$	&tet	&4&--\\
$\beta$-Nb		& -0&46	&\bf	0&\bf00	&\multicolumn{2}{c}{}	&$12d$	&tet	&--&4\\
\midrule
ZrFe$_2$$^\dag$	& -0&03	&   	0&50		& $\pm0$&00 		&$96g/6h$&tet	&2&2\\
ZrMo$_2$			& -0&24	&  	0&22		& $\pm0$&06 		&$96g/6h$&tet	&2&2\\
ZrCr$_2$$^\dag$	& -0&31	& 	0&15		& $\pm0$&03  		&$96g/6h$&tet	&2&2\\
ZrV$_2$			& -0&73	&\bf-0&\bf26	& $\pm0$&01 		&$96g/6h$&tet	&2&2\\
ZrNb$_2$			& -0&81	&\bf-0&\bf35	& $\pm0$&02		&$96g/6h$&tet	&2&2\\
\midrule
ZrSn$_2$			&   0&28	&   	0&75		&\multicolumn{2}{c}{}	&$32h$	&tet	&2&2\\
ZrCu				& -0&28	& 	0&19		&\multicolumn{2}{c}{}	&$3d$	&oct	&4&2\\
ZrNi				& -0&37	&  	0&09		&\multicolumn{2}{c}{}	&$4c$	&oct	&4&2\\
Zr$_5$Sn$_{3+x}^{\ast\ast}$
				& -0&38	&\bf 0&\bf08	& $\pm0$&14		&$2a/2b$	&tri/oct&3--6&--\\
Zr$_2$Fe$^\dag$	& -0&45	&\bf-0&\bf01	&\multicolumn{2}{c}{}	&$16l$	&tet	&4&--\\
Zr$_2$Cu			& -0&52	&\bf-0&\bf06	&\multicolumn{2}{c}{}	&$4d$	&tet	&4&--\\
Zr$_2$Ni	$^\dag$	& -0&67	&\bf-0&\bf20	&\multicolumn{2}{c}{}	&$16l$	&tet	&4&--\\
Zr$_3$Sn			& -0&64	&\bf-0&\bf18	&\multicolumn{2}{c}{}	&$6d$	&tet	&4&--\\
Zr$_3$Fe			& -0&74	&\bf-0&\bf27	&\multicolumn{2}{c}{}	&$8f$	&tet	&4&--\\
\bottomrule
\end{tabular}


\end{table}
In the ZrM$_2$ phases, little variation in the accommodation of H was observed between the three Laves structures $C14, C15 \text{ and } C36$, therefore the results from ZrM$_2$ simulations have been condensed into a single value for each element and the statistical error is reported.
Similarly, the three models of the Zr$_5$Sn$_{3+x}$ phase were reported as one. The exact values of $E_{sol}$(H) were $-0.56$, $-0.34$, $0.23$ eV  for $x = 0.0, 0.5, 1.0$ respectively. For the case of Zr$_5$Sn$_4$, the lowest energy site was the only unoccupied $2b$ Wyckoff site. It is reasonable to expect a large number of unoccupied Sn self-interstitial sites in this phase, therefore the defect energy was calculated as an interstitial defect into a phase with one unoccupied Sn self-interstitial site, rather than the substitutional defect H$_\text{Sn}$ in the fully occupied Zr$_5$Sn$_4$ structure.
Zr$_5$Sn$_{3+x}$, exhibits two preferred sites: the $2a$ site between 3 Zr atoms, and, directly above and below it, the $2b$ site between 6 Zr atoms.
The relative preference for H of one site over the other changes as a function of Sn content. At a content of three Sn atoms per formula unit, the trigonal $2a$ site is preferred (\SI{-0.56}{eV} against \SI{-0.39}{eV} of the octahedral site). As the Sn content increase to 3.5, half of the $2b$ sites are occupied by the excess Sn. The presence of a Sn atom in the $2b$ site, causes a reduction in space in the neighbouring $2a$ sites, while only marginally affecting the configuration of other (unoccupied) $2b$ sites in the cell. This is reflected in the solution of H in the two sites: the $2a$ site becomes significantly less favourable (\SI{-0.27}{eV}), while the $2b$ provides a similar solution enthalpy as in the previous case (\SI{-0.34}{eV}) . In the case of Zr$_5$Sn$_4$ all of the $2b$ sites are occupied, consequently the $2a$ sites are compressed by two Sn atoms, one above and one below, reducing the volume available for accommodation of H even further, and the enthalpy associated with accommodating an H atom in that site becomes positive and large (\SI{2.06}{eV}).

The current work shows that $\beta$-Zr accommodates H more readily than $\alpha$-Zr (in agreement with experimental data\cite{Ells1956}), and that $\beta$-Nb exhibits the same value of $E_{sol}$(H) as $\alpha$-Zr. This suggests that, if the $\beta$-(Nb,Zr) solid solution found in binary Zr-Nb alloys behaves similarly to its two end members, those alloys do not contain any strong sinks for H.
Nevertheless, the formation of metastable ZrNb$_2$ phases may affect this, as discussed below.

Regarding the ZrM$_2$ Laves phases, the solution enthalpy of H generally decreases with increasing number of $d$ electrons in the transition metal M: from highest to lowest affinity Nb, V, Cr, Mo  and Fe. The same trend has been observed with respect to H solution capacity\cite{Shaltiel1977}.
Whilst H does not prefer to dissolve in the Laves phases containing the latter three elements (i.e. Cr, Mo, and Fe) compared to $\alpha$-Zr, the intermetallics formed with either Nb or V offer favourable sites for the accommodation of H.
This suggests that if these binary SPPs are present in the cladding, H will likely segregate to them, which may deplete the H content in the zirconium metal.
The beneficial effect of the H sinks may, however, be limited to the initial stages of the fuel cycle. At higher burnups, the intermetallic particles are likely to dissolve\cite{Griffiths1987}, amorphise or oxidise, thereby releasing any stored H. 

\begin{figure}[p]
\centering
\includegraphics[width = 0.45\textwidth]{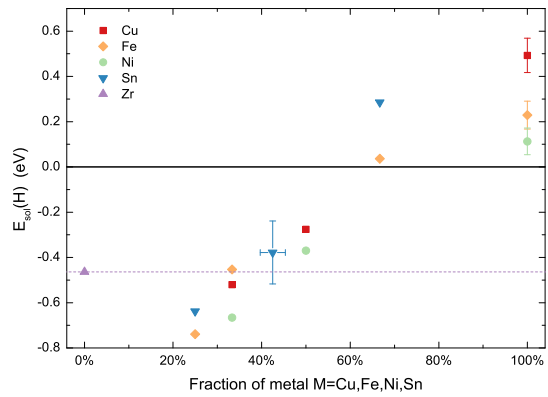}
\caption{\label{fig:H_sol-M_content} Enthalpy of H solution {\it vs.}~fraction of non-Zr metal in compound. Values for the pure metals M were taken from experimental and previous DFT work\cite{McLellan1975,Aydin2012}. No published results are available for H in Sn.}
\end{figure}

In addition to ZrNb$_2$ and ZrV$_2$, all Zr-rich phases provided lower $E_{sol}$(H) values compared to $\alpha$-Zr.
Furthermore, for each element where more than one stoichiometric phase is present (Cu, Fe, Ni and Sn), those with the largest Zr/M ratio provided the lowest solution enthalpy for H (see Figure~\ref{fig:H_sol-M_content}).
Zr is known to exhibit higher affinity for H compared to Cu, Fe and Ni (see the extremes of Figure~\ref{fig:H_sol-M_content}), therefore a decrease in enthalpy of solution with increasing Zr content is expected. However, from a volumetric standpoint, intermetallic phases have a lower packing fraction compared to the pure metals, offering a larger number of interstitial sites with varying amounts of space. For this reason intermetallic phases are expected to exhibit lower defect volumes (and associated strain fields) when accommodating an H interstice.
As a result of the two competing processes --- chemical bonding and volumetric effects --- the lowest solution enthalpies are found, as mentioned above, for Zr-rich intermetallic phases that provide interstitial sites with 4-fold Zr-coordination.

Unlike other alloying additions, Fe forms a wide range of intermetallic compounds, and their relative stability is greatly affected by other alloying elements: Zr$_2$(Fe,Ni) SPPs are commonly observed in the absence of Cr additions, while in Cr-containing alloys, Zr(Cr,Fe)$_2$ Laves phases become the dominant SPPs\cite{Barberis2005}.
In the presence of Nb, hexagonal Zr(Nb,Fe)$_2$ Laves phases and cubic (Zr,Nb)$_2$Fe phases have been reported\cite{Barberis2004}.
Similarly, the few records of Sn-Zr SPPs also report some Fe solubility into these particles\cite{Griffiths1987,Griffiths1988,Yang1987}.
This suggests that the addition of Fe does not influence which intermetallic phases form, rather it will go into solution in all or most of the SPPs present.
Assuming that a ternary phase behaves similarly to its binary end members, solution of Fe in ZrM$_2$ Laves phases and Zr$_2$X phases, is expected to reduce their affinity to H, since ZrFe$_2$ and Zr$_2$Fe exhibit less favourable $\Delta E_{sol}(\text{H})$ compared to all other phases with the same stoichiometry and structure. The opposite can be said for the Zr$_3$Fe-type SPPs.

Owing to the fact that both ZrCr$_2$ and ZrFe$_2$ have unfavourable $\Delta E_{sol}(\text{H})$ values, it is predicted that the ternary Zr(Cr,Fe)$_2$ phase, found predominantly in Zircaloy-4, does not getter H from the surrounding $\alpha$-Zr. As for the Zr$_2$(Fe,Ni), which is the predominant SPP in Zircaloy-2, both of its binary end members offer favourable solution enthalpies for H. Nonetheless, the difference in affinity to H with respect to $\alpha$-Zr is rather small (on the order of \SI{0.1}{eV}), and is expected to diminish with increasing Fe content. Such small differences in energy may easily be overcome by thermal and entropic effects, consequently Zr$_2$(Fe,Ni) are not expected to be strong sinks for H in solution. 

Whilst it is possible to speculate on the behaviour of ternary phases where both binary end members have either positive or negative $\Delta E_{sol}(\text{H})$ values, it is much harder to predict the H affinity of other ternary SPPs such as Zr(Nb,Fe)$_2$ found in ZIRLO alloys. It is possible, however, to expect that the behaviour of such SPPs is strongly correlated to their chemical composition, and more specifically the Nb/Fe ratio of the intermetallic particle. If $\tfrac{\text{Nb}}{\text{Fe}} >> 1$ then the SPPs may act as H sinks, whilst with a composition of $\tfrac{\text{Nb}}{\text{Fe}} << 1$ they are not likely to accommodate any H.

Under irradiation, Fe has been reported to diffuse out of the SPPs faster than other elements\cite{Yang1986,Griffiths1987}. In the case of Laves phases, and especially Nb-containing Laves phases, the current work suggests that this would increase the affinity of the residual SPP for H. However, concomitantly to the dissolution of Fe, the SPPs are reported to undergo amorphisation and at present it is impossible to predict how this will affect the interaction between H and the SPP.

\section{Conclusions}
DFT simulations have been employed to calculate the enthalpy of solution of H in pure Zr, Nb and in intermetallic phases of Zr.
Pure $\beta$-Zr exhibits more favourable solution enthalpy for H relative to $\alpha$-Zr, however the difference is predicted to diminish in the presence of Nb.
Regarding the Laves SPPs, the presence of Nb or V is predicted to increase the affinity with H, whilst the presence of Cr, Mo and Fe will reduce it.
Cu, Ni and Sn additions may form a number of binary intermetallic phases, but tend to stabilise as Zr-rich phases.
All Zr-rich phases, namely Zr$_3$Fe, Zr$_2$Ni, Zr$_2$Cu and Zr$_3$Sn, provide lower energy sites for H accommodation, compared to $\alpha$-Zr, suggesting that their presence in the alloy could provide sinks for H.

With regards to the more commonly observed ternary SPPs, it is predicted that Zr$_2$(Fe,Ni) (found mainly in Zircaloy-2) and $\beta$-(Zr,Nb) precipitates (present in all Nb containing alloys) exhibit an affinity to H similar to that of $\alpha$-Zr, and are therefore not expected to strongly getter H from their surroundings.
Zr(Cr,Fe)$_2$ SPPs found in Zircaloy-4 are predicted to have very unfavourable solution enthalpies for H and therefore not to accommodate any H. The affinity to H of the Zr(Nb,Fe)$_2$ SPPs, present in ZIRLO alloys, is expected to vary strongly with Nb/Fe ratio: high Nb content SPPs are expected to trap H, whilst high Fe content SPPs will reject it.


\section{Acknowledgements}
We would like to thank the EPSRC, ANSTO and the UK-MOD for financial support. We also acknowledge Imperial College HPC for the use of resources.
M.R.\ Wenman acknowledges support from EDF Energy through a Fellowship scheme.





\section{References}








•

•


\end{document}